\documentclass{article}

\usepackage{gensymb} % For \degree symbol                                                                            
\usepackage{graphicx}%
\usepackage{amsthm}%                                                                                    
\usepackage[title]{appendix}%                                                                           
\usepackage{xcolor}%
\usepackage{pdfpages}
\usepackage{cite}
\usepackage[T1]{fontenc}
\usepackage[utf8]{inputenc}
\usepackage{amsmath,amsfonts,amssymb}
\usepackage{graphicx}
\usepackage{xcolor}
\usepackage{manyfoot}
\usepackage{xr-hyper}
\usepackage{hyperref}
\usepackage{braket}
\usepackage{geometry}

\begin{document}

\title{Topological invariants and topological charges in photonic systems}

\author{Kristian Arjas$^1$, Grazia Salerno$^{1,2}$, Päivi Törmä$^{1,\dagger}$}
\date{
$^1$Department of Applied Physics, Aalto University School of Science, P.O. Box 15100, Aalto FI-00076, Finland\\
$^2$Dipartimento di Fisica ``E. Fermi'', Università di Pisa, Largo Bruno Pontecorvo3, Pisa, 56127, Italy\\
$^\dagger$ Corresponding author(s). E-mail(s): paivi.torma@aalto.fi
}

\maketitle

\begin{abstract}
Topology in photonics comes in two distinct flavors: global and local.
Global topology considers invariants that are obtained by integrating over the energy band, whereas local topology considers defects, typically vortices, in the far-field emission.
These topologies are described by a wide range of models built in both real and momentum space, which are connected only by computationally expensive numerical methods that lack physical intuition.
Here we propose a general framework based on a real-space Hamiltonian capable of describing electric field as a vector in both near- and far fields, allowing us to bridge between topological defects in the far-field and global topological invariants.
The proposed Hamiltonian is constructed from the symmetry-representations of the lattice, is deformable to both atomic localized-mode (tight-binding) and photonic delocalized-mode (long-range) limits, and allows for independent control over the energies of eigenmodes of different symmetries at high-symmetry points of the Brillouin zone.
This symmetry-based approach enables the design of structures with almost arbitrary topological properties and is not limited to photonic systems, but could apply to any system with engineered real-space couplings.
\end{abstract}

\section*{Introduction}

Topological photonics promises unidirectional edge states that hold great potential for the robustness of the next generation of photonic technologies.
Ever since the inception of the field by Haldane and Raghu \cite{haldane2008possible,raghu2008analogs}, a wide variety of topologically non-trivial states have been realized~ \cite{rider2019perspective,ozawa2019topological,kang2023topological,liu2021topological}.
In many cases, the bands are designed to carry a global topological invariant, such as the Chern number or Zak phase, which are obtained by integrating the Berry curvature or the Berry connection over the whole Brillouin zone (BZ).
Due to weak magneto-optical effects at optical frequencies, alternative topological invariants have also been studied extensively, such as inversion-broken systems with Valley-Chern numbers \cite{ma2016all,dong2017valley} and $\mathbb{Z}_2$ invariants associated with pseudospin~\cite{wu2015scheme}, or both \cite{chen2020coexistence}.

In photonic systems, topological defects have appeared as an additional line of research alongside the renowned topological invariants. 
These defects are associated with bound states in the continuum (BICs), special nonradiative solutions of the wave equation \cite{hsu2016bound}, associated with polarization vortices of non-zero topological charges, i.e., winding numbers in momentum space.
BICs have been experimentally realized in a wide range of electromagnetic systems, including photonic crystals and plasmonics, and have enabled lasing from with topological charges ranging from low (1-2) in periodic systems~\cite{Miyai2006,iwahashi_higher-order_2011,KodigalaLasingBIC,LasingHa,huang2020ultrafast,guan2020quantum,LasingWu,wangTopologicalCharge2020,hwang2021ultralow,wu2021,heilmann2022quasi,Hakala,salerno2022loss,Sang2022,ren2022,zhai2023,berghuis2023room, do2025room, xue2024lasing, song2025high, eyvazi2025flat} to high (5-19) in quasiperiodic systems \cite{arjas2024high}.
The topology exhibited by these polarization vortices is connected to a geometric phase \cite{bliokh2019geometric}, a feature that has been exploited when converting spin-polarized light to optical vortex beams \cite{wang2020generating}.

Topological quantities can be extracted by solving the photonic bands of the system, typically by numerically solving Maxwell's equations in the frequency domain using, e.g., finite element and finite difference time domain solvers~\cite{johnson2001block,minkov2020inverse}.
While these methods are powerful, they lack the intuition given by models based on a Hamiltonian description. Depending on the dominant coupling scale, these models can either be in the tight-binding (TB) limit~\cite{kang2023topological,amo2016exciton,pocock2018topological,ling2015topological,weick2013dirac,rechtsman2013photonic,FloquetPlasmonic,li2023polarization, schurr2025plasmonic,liu2017novel,xie2018second}, or described by the empty-lattice approximation (ELA), which based on diffracted orders~\cite{wang2018rich,freire2025plasmonic, schokker2016lasing, ramezani2016plasmon, guo2019lasing, de2018interaction, hakala2018bose}.
Connecting the topological properties of TB and ELA models is non-trivial due to the fundamentally different nature of short-ranged versus long-ranged couplings, respectively. 

Some work has been done in extending TB Hamiltonians to consider either longer distance couplings \cite{chen2020elementary, hoang2024photonic} or larger unit-cells \cite{ma2022electronic}.
Here we propose a generic framework capable of modeling the electric field properties of photonic lattices using real-space Hamiltonians with long-range couplings.
These couplings are not free parameters, but are determined by the inverse Fourier transform of the desired band structure, allowing us to connect the TB models with the diffractively coupled ELA models.
We introduce a group-theory-based method that systematically treats the symmetry of the unit cells and enables independent control of the mode energies at high-symmetry points (HSPs) of the BZ.
Our methodology is extended to vector fields, where each site in the unit cell is treated as a dipole, creating a natural link between polarization vortices in the near and far fields.
The control over the mode orderings at HSPs allows us to use topological quantum chemistry~\cite{bradlyn2017topological,morales2025transversality} as a design tool for the realization of topologically nontrivial bands, the study of which can be taken further with standard Hamiltonian methods.

Because our model is built from symmetry principles, its momentum-space formulation is analytical, and the corresponding real-space Hamiltonian can be readily obtained via the inverse Fourier transform. 
Although the resulting real-space model is not immediately tied to a specific dielectric function profile of a photonic crystal, it provides valuable insight into the spatial couplings between sites in engineered structures. 
The real-space formulation provides a powerful framework for the experimental design of topological lattices in many synthetic engineered materials, even beyond photonics, including circuit lattices or mechanical metamaterials that allow for direct design of site-to-site couplings~\cite{shah2024colloquium, lee2018topolectrical, wang2015topological, nash2015topological}.

For the vector field extension of the model, we construct bands with different topological invariants and observe how the far-field emission adapts to changes of the HSPs modes.
For bands carrying Chern numbers, we compare the Berry curvature calculated from the eigenvector with the quantity one would observe in far-field radiation (so-called radiation curvature~\cite{yin2025observation, yuan2025breakdown}).
By constructing structures with open boundary conditions, we evaluate the properties of the topological modes in the far field. The edge state and its far field emission remain robust under large structural deformations, highlighting the power of our framework in understanding, analyzing, and designing protected topological photonic systems.

\subsection*{Constructing the Hamiltonian}
We present the construction of the effective Hamiltonian, consisting of two parts: 
\begin{equation}
\hat{H}(\mathbf{k}) = \hat{H}_k(\mathbf{k}) + \hat{H}_c . 
\end{equation}
The first term $\hat{H}_k(\mathbf{k})$ is a momentum ($\mathbf{k}$) dependent part that approximates the ELA, while the second part $\hat{H}_c$ is a momentum-independent coupling matrix that shifts the energies of the bands associated with different symmetry modes at HSPs.

\subsubsection*{Scalar fields}
We begin by considering a simplified case where the photonic mode can be treated as a scalar field, appropriate for describing, e.g., the out-of-plane electric or magnetic field components in systems with horizontal symmetry. 

The momentum space Hamiltonian is constructed in the absence of any coupling beyond the crystalline symmetry constraints, thus it captures the dispersion of the freely propagating modes $\hat{H}_k(\mathbf{k}) = \sum_\mathbf{k} \varepsilon(\mathbf{k}) \hat a_\mathbf{k}^\dagger \hat a^{}_\mathbf{k}$. We require the dispersion $\varepsilon(\mathbf{k})$ to be written as $\varepsilon(\mathbf{k}) = \sum_\mathbf{m} t_\mathbf{m} e^{i \mathbf{k} \cdot \mathbf{m}}$, where $t_\mathbf{m}$ is the Fourier coefficients of the desired ELA. By performing an inverse Fourier transform $\hat{a}_\mathbf{k}^\dagger = \sqrt{\frac{N}{2\pi}} \sum_n e^{-i \mathbf{k} \cdot \mathbf{n}} \hat{a}_\mathbf{n}^\dagger$, the following Hamiltonian is obtained
\begin{align*}
    \hat{H}_k = \sum_\mathbf{m,n} t_\mathbf{m} \hat a_\mathbf{n}^\dagger \hat a^{}_{\mathbf{m}-\mathbf{n}}
\end{align*}
which represents a real-space model where two sites, one located at position $\mathbf{n}=(n_x, n_y)$ and another one whose distance from the first is $\mathbf{m}= (m_x, m_y)$, are coupled. Generally, TB models consider only nearest-neighbor couplings, and $|\mathbf{m}| = \sqrt{m_x^2+m_y^2}=1$.

Additional next-to-nearest neighbor couplings are needed to reproduce the 2D ELA dispersion relation, which is the light cone dispersion in the first BZ: $\varepsilon(k_x,k_y) = \frac{\hbar c}{n}\sqrt{k_x^2 + k_y^2}$ with $k_x,k_y \in [-\pi,\pi]$ and $n$ being the refractive index of the medium.
The couplings $t_\mathbf{m}$ needed to approximate the light-cone dispersion are obtained with a multi-index $\mathbf{m} : \{m_x,m_y\}$, $m_{x,y} = 0,\pm1,\dots\pm M$, up to an arbitrary range $M$.

\begin{figure*}
    \centering    
    \includegraphics[width=0.95\textwidth]{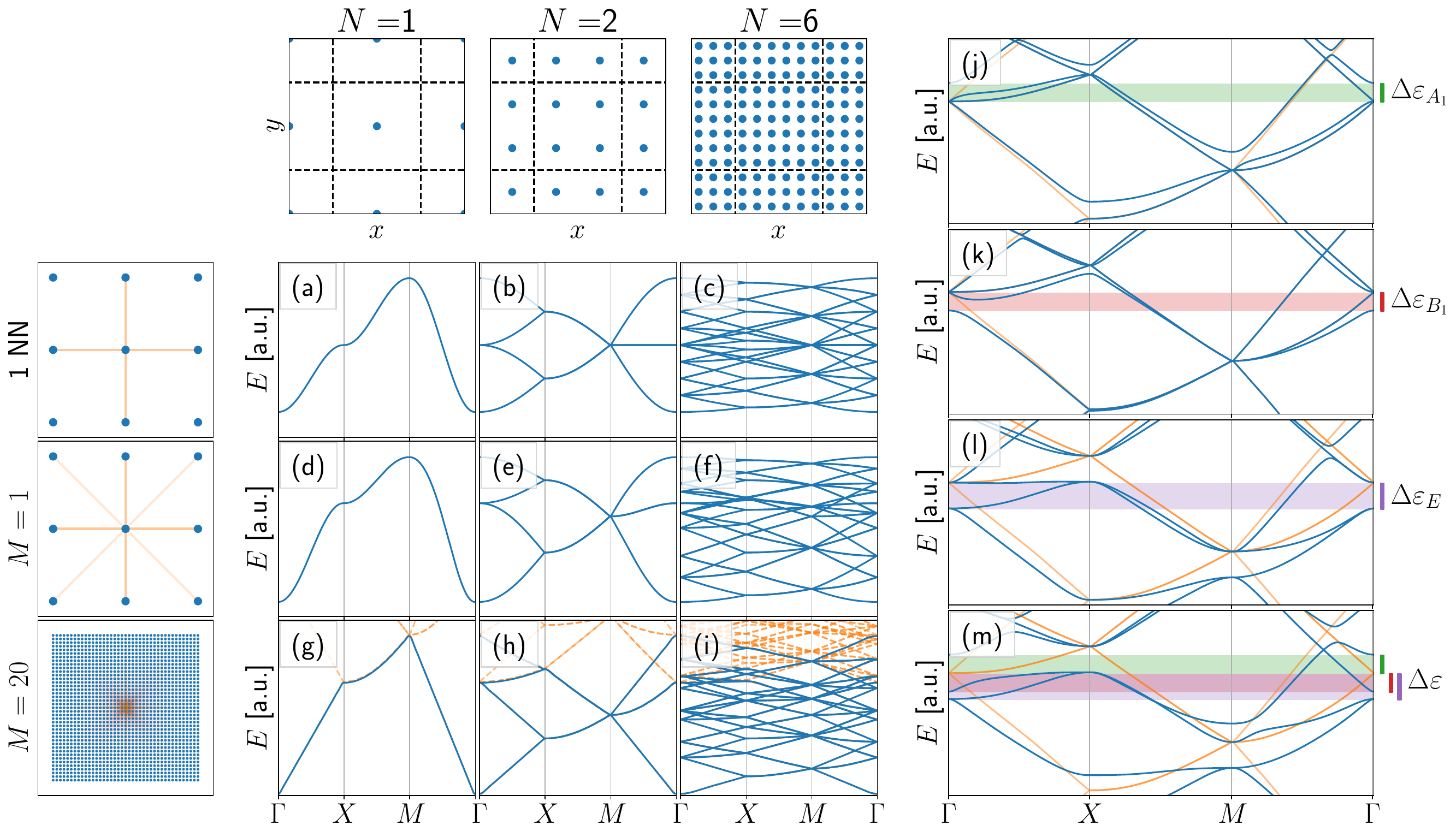}
    \caption{
        The evolution of the band-dispersion Hamiltonian $\hat H_k$ (a-i).
        The band-dispersion evolves from cosine to linear with the inclusion of more coupled neighbors (rows), up to $M = 20$.
        On the third row, the ELA dispersions are plotted in orange (they are also under the blue lines).
        Folding the Brillouin zone by enlarging the unit cell allows for more bands (columns).
        The coupling Hamiltonian $\hat{H}_c$ allows us to target the energies of the different symmetry eigenmodes independently (j-m).
        Introducing a coupling of $\Delta \varepsilon_{IR}P_{IR}$ shifts their respective IRs by $\Delta \varepsilon_{IR}$.
        (m) combines the couplings from (j-l).
    }
    \label{fig:model_evolution}
\end{figure*}

The effect of introducing next-to-nearest neighbor couplings is shown in Fig.~\ref{fig:model_evolution} (a), (d), and (g). 
As more neighbors are progressively coupled, the band evolves from the usual TB dispersion $\varepsilon(k_x, k_y) = 2t \left[\cos k_x + \cos k_y\right]$ in Fig.~\ref{fig:model_evolution}(a), to the desired linear ELA in Fig.~\ref{fig:model_evolution}(g).

Higher energy bands can be accessed by enlarging the unit cell with $N_{x,y}$ sites along the x, y direction, so that the analytical form of the Hamiltonian changes as $H_k(k_x, k_y) \rightarrow H_k(k_x/N_x, k_y/N_y) = H_k(k'_x, k'_y)$. The BZ is thus folded, as shown in Fig.~\ref{fig:model_evolution} (a-i) for systems with increasingly longer-ranged couplings.
The general form of $\hat H_k$ is given in the Methods.

Due to the homogeneity of $\hat{H}_k$, gaps between bands are only opened by the momentum-independent coupling Hamiltonian $\hat H_c$.
We consider a symmetry group $\mathcal{G}$ compatible with the lattice of interest.
The group consists of elements (symmetry operations) $g_i\in \mathcal{G}$ which have representation matrices $D(g_i)$ (symmetry operators).
These representations can be broken down into irreducible representations (IRs), which are the eigenbasis of the symmetry group.
Since the Hamiltonian commutes with the group symmetries, each eigenvector of the Hamiltonian belongs to one of the IRs.
Thus, each IR consists of sets of eigenvectors that transform similarly under symmetries in the group.

We construct projection matrices $P_\text{IR}$ for each IR as~\cite{inui2012group}
\begin{equation}
    P_\text{IR} = \frac{1}{g}\sum_i\epsilon_{i,\text{IR}}D(g_i),
    \label{eq:projections}
\end{equation}
where $g$ is the size of the group, $i$ sums over all symmetry elements and $\epsilon_{i,\text{IR}}$ is the character of the IR for symmetry-operation $i$, tabulated in the character tables.

By using these matrices, we project the couplings onto the Hamiltonian in such a way that it only affects a single IR.
Due to their orthogonality, the IRs can be targeted independently.
The full coupling Hamiltonian is then constructed as
\begin{equation}
    \hat{H}_c = \sum_{i\in \text{IR}} \Delta \varepsilon_i P_i,
    \label{eq:H_c}
\end{equation}
which has the effect of opening gaps and ordering the energies of the IRs at HSPs based on the values of $\Delta \varepsilon_i$.
This concept is demonstrated in Fig.~\ref{fig:model_evolution} (j-m).

If a degeneracy contains multiple modes in the same IR, they cannot be separated by a constant $\Delta\varepsilon_\text{IR}$.
In those cases, additional couplings can be introduced in the system, which can be projected onto a specific IR with $P_{IR}$.

\subsubsection*{Vector fields}

\begin{figure*}
    \centering
    \includegraphics[width=0.95\textwidth]{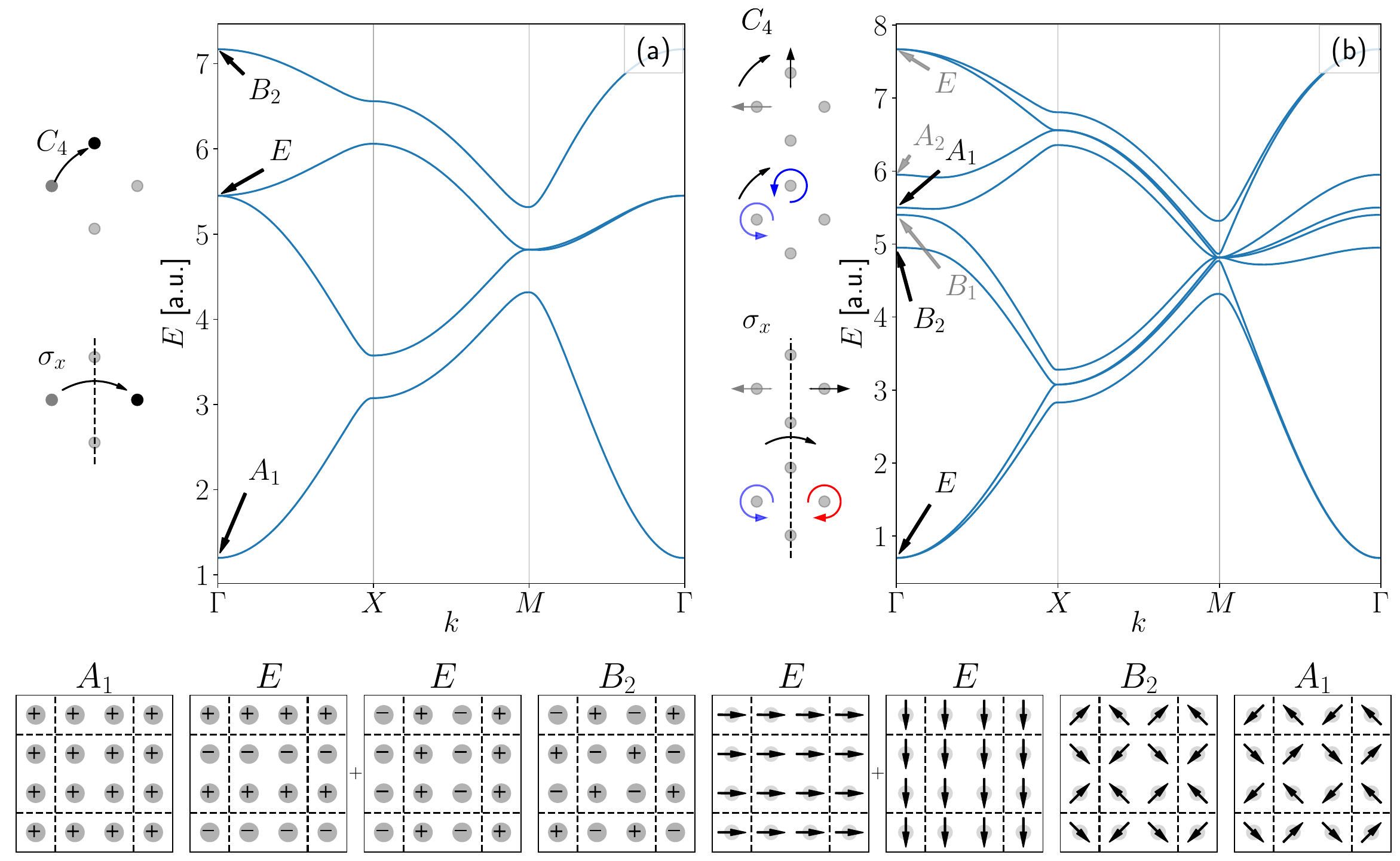}
    \caption{
    (a) A scalar 2x2 system with $M = 1$ couplings (panel (e) in Fig.~\ref{fig:model_evolution}) and $\Delta \varepsilon_{A} = 1/2, \Delta \varepsilon_{B} = -1/2$ with the symmetry operations on the unit-cell depicted on the left.
    The four eigenvectors are depicted in the bottom panels.
    (b) The vectorized model with the same IR couplings, with four eigenvectors displayed, highlighted in the band dispersion with the effect of symmetry operations shown for both dipoles and circular components.
    The $B_1$ and $A_2$ are 90-degree rotations of the modes $B_2$ and $A_1$ shown here, respectively. 
    }
    \label{fig:vectorization}
\end{figure*}
So far, we have described a scalar field, i.e, one degree of freedom per site.
However, the in-plane electric field is a vector quantity with $x$ and $y$ components, thus a description of photonic lattices requires extending the model.
Here we use the circular polarization basis, namely $\ket{L} = \ket{x} + i\ket{y}$ and $\ket{R} = \ket{x} - i\ket{y}$ as our two degrees of freedom.
Since $\ket{L}$ and $\ket{R}$ are orthogonal, $\hat H_k(\mathbf{k})$ should be identical for the two.
The expansion of $\hat H_k(\mathbf{k})$ is straightforwardly done as
\begin{equation}
    \hat H_k(\mathbf{k}) \rightarrow I_2\otimes \hat H_k(\mathbf{k}),
    \label{eq:Hk_vec}
\end{equation}
which expands our basis from $\ket{ij}\rightarrow \ket{LR}\otimes \ket{ij}$.

The coupling Hamiltonian $\hat{H}_c$ for a vector field, however, needs to be modified because the vector nature of the field has to be taken into account when constructing the representation matrices $D(g_i)$ in Eq.~\eqref{eq:projections}.
In both vector and scalar cases, the effect of symmetry operations on the sites $\ket{ij}$ remains the same.
For vectors, we need additional representation matrices $D'(g_i)$ that act on the $\ket{LR}$ basis, see Methods.
Combining these operators, we construct the vectorial representation matrices $D_v(g_i)$ as 
\begin{equation}
    D_v(g_i) = D'(g_i)\otimes D(g_i),
    \label{eq:repr_vec}
\end{equation}
which act on our basis as $D_v(g_i)\ket{LR,ij} = D'(g_i)\ket{LR}\otimes D(g_i)\ket{ij}$.
Thus, the projection matrices are
\begin{equation}
    P_\text{IR} = \frac{1}{g}\sum_i\epsilon_{i,\text{IR}}D_v(g_i).
    \label{eq:proj_vec}
\end{equation}
Figure~\ref{fig:vectorization} shows the change of the representations of the IRs in the vectorial case as opposed to the scalar case.

\subsubsection*{Far fields} 
The eigenvectors $\ket{\psi(\mathbf{k})}$ of the total Hamiltonian, defined by $\left(\hat{H}_k(\mathbf{k}) + \hat{H}_c \right) \ket{\psi(\mathbf{k})} = \varepsilon(
\mathbf{k})\ket{\psi(\mathbf{k})}$, define the fields in momentum space. 
These are the fields inside and in the immediate vicinity (evanescent) of the photonic structure. In the following, we refer to these as near fields, to distinguish from the field radiated far away from the system. 

The far fields are obtained as a coherent sum of the eigenstates on the sites of the unit cell in the near field, as
\begin{equation*}
    \ket{E(\mathbf{k})} = Q \ket{\psi(\mathbf{k})},
\end{equation*}
where $Q = I_2\otimes \ket{1}_N$, with $\ket{1}_N$ being a row vector of $N$ ones~\cite{yin2025observation, yuan2025breakdown}.
For example, for a state $\ket{\psi(\mathbf{k})} = \left(L_1, \dots, L_N, R_1, \dots, R_N \right)^T$ defined on a $N$-sites unit cell, the far field is $\ket{E(\mathbf{k})} = \left(\sum_i^N L_i, \sum_i^N R_i\right)^T$.

\begin{figure}
    \centering
    \includegraphics[width=0.5\linewidth]{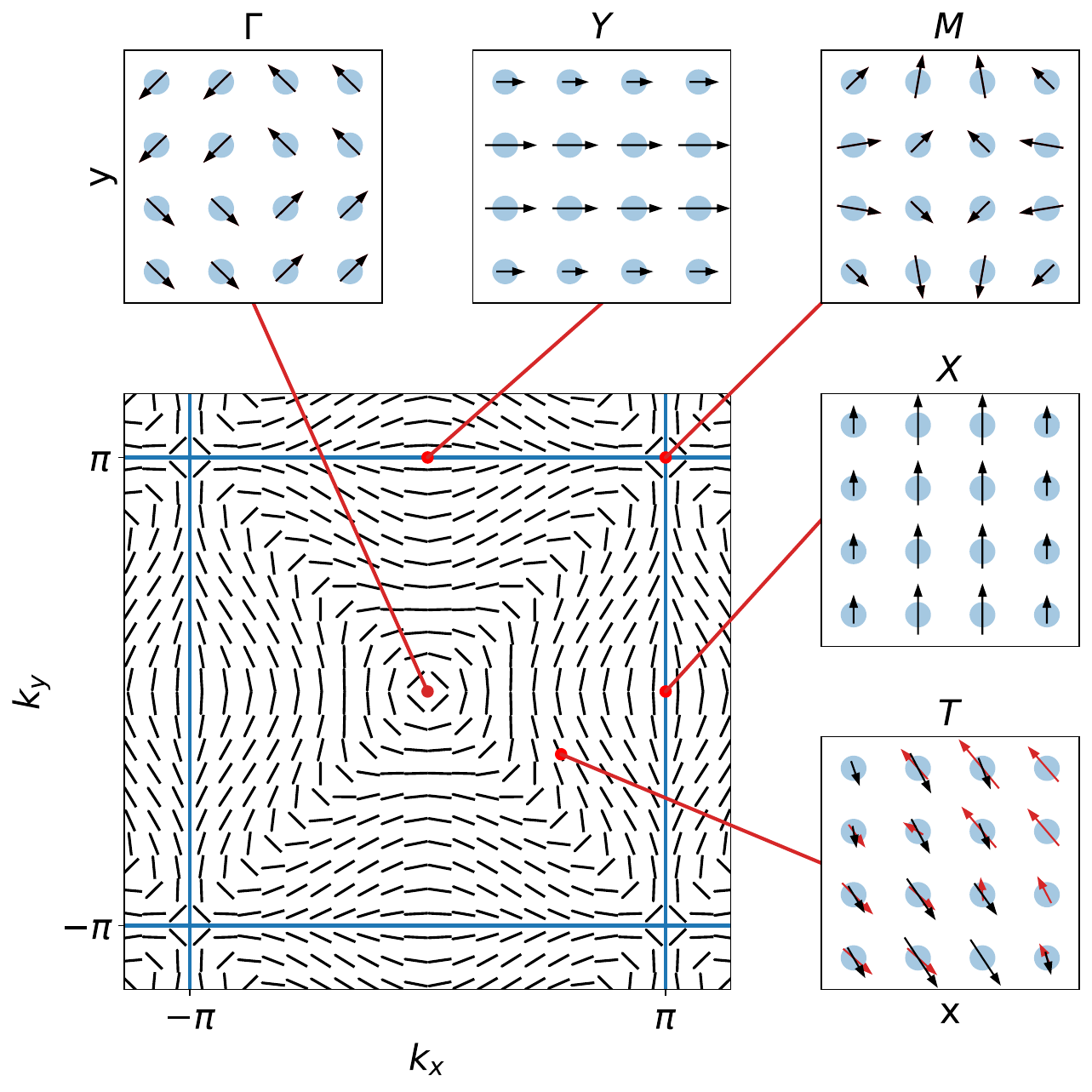}
    \caption{
    The connection between polarization ellipses (which appear flat due to the lack of circular polarization) in the far field and the near field at the HSPs of the BZ and a trivial point $T$.
    The black and red arrows show the real and imaginary parts of the electric field.
    At $\Gamma$ and $M$-points, the net field can be seen to vanish.
    It should be noted that the symmetry properties are the same in near- and far-field.
    }
    \label{fig:nf_ff}
\end{figure}

Since $Q$ is not a square matrix, it is not invertible.
Hence, $\ket{E}$ is uniquely defined by $\ket{\psi(\mathbf{k})}$ but not the other way around.
The polarization properties in the far field are calculated from $\ket{E(\mathbf{k})}$ via the Stokes parameters $S_i(\mathbf{k}) = \bra{E(\mathbf{k})} \hat{\sigma_i} \ket{E(\mathbf{k})}$, where $\sigma_i$ is the i-th Pauli matrix.
The connection between near- and far fields is depicted in Figure \ref{fig:nf_ff}.

\subsection*{Band topology}
Topological band theory defines topological invariants, including Chern numbers $C$ or Zak phases $Z$, by integrating geometrical quantities of the Hamiltonian, such as the Berry curvature, over the full BZ. These invariants are not sensitive to deformations of the system's parameters that leave the gaps to other bands open.

In our system, we can calculate two kinds of curvatures. The first is the conventional Berry curvature defined from the eigenvectors of the Hamiltonian $\ket{\psi}$ (the near field) as
\begin{equation}
    B_\text{NF}(k_x,k_y) =  i \left(\langle \partial_{k_x} \psi | \partial_{k_y} \psi\rangle -\langle \partial_{k_y} \psi | \partial_{k_x} \psi\rangle\right) .
\end{equation} 
The second is the radiation curvature calculated from the far-field, which take the following form~\cite{fosel2017lines, bleu_measuring_2018, hu2024generalized}   
\begin{equation}
    B_\text{FF}(k_x,k_y) = \frac 1 2 \sin\chi \left(\partial_{k_x}\chi \partial_{k_y}\phi - \partial_{k_y}\chi \partial_{k_x}\phi\right) ,
    \label{eq:berryff}
\end{equation}
where the radial and azimuthal polarization angles $\phi = \frac{1}{2}\arg(S_1(\mathbf{k}) + iS_2(\mathbf{k}))$ and $\chi= \arccos(S_3(\mathbf{k}))$ are defined by the Stokes parameters. 
From the $\sin\chi$ in Eq.~\eqref{eq:berryff}, one can conclude that the far-field Berry curvature is most sensitive to variations in polarization when  $\chi \approx 0$ or $\chi \approx \pi$, i.e., around purely left- or right circularly polarized points.

The Chern number is obtained by integrating the Berry curvature over the BZ. Chern numbers appear in systems with broken TRS, and their value corresponds to the number of unidirectional edge states. 
%As a $\mathbb{Z}_2$ topological invariant, Zak phases instead can be derived from the symmetry properties of the eigenvectors at $\Gamma$ and $X$-points, which we utilize in the following.
It has been shown that the symmetry-eigenvalues at the HSPs of a band can fully predict the Zak phase, while the Chern number is predicted modulo the degree of rotational symmetry \cite{vaidya2023topological,fang2012bulk}.

\subsection*{2D-SSH-like models}

We now apply our formulation to build on the 2D Su-Schrieffer-Heeger (2D-SSH) model in a controlled fashion. 
We recover the standard 2D SSH Hamiltonian~\cite{liu2017novel} for system of size $N_x = N_y = 2$ and couplings $t_\mathbf{m} = t\cdot\delta_{|\mathbf{m}|^2, 1}$ with $t < 0$. 
The corresponding dispersion is shown in Fig. \ref{fig:model_evolution}(b).
The difference between inter- and intracell couplings $t$ and $t_1 = t + \delta$ is created by $\hat H_c$ by using the $\mathbf{C}_4$ representation matrices with $\Delta\varepsilon_{A} = -\Delta\varepsilon_B = \delta/2$.
The model is topological when $\delta > 0$.
The matrices are explicitly given in the Methods.

Magnetic effects can be introduced to the model by lifting degeneracy between $E_1$ and $E_2$.
For simplicity, we set $\Delta m/2 = \Delta \varepsilon_{E_1} = -\Delta\varepsilon_{E_2}$ to make the intracell couplings complex.
Now $t_1 = |t'| e^{i \varphi_m}$ with a Peierls phase $\varphi_m =  \arctan\left[\Delta_m/(t+\delta)\right]$ and $|t'| = \sqrt{(t+\delta)^2+\Delta_m^2}$, while the intercell couplings $t$ remain real. 
Such a situation breaks time reversal symmetry (TRS) and corresponds to a flux lattice, where alternating plaquettes have magnetic fluxes of $4\varphi_m$ and $-2\varphi_m$, see Methods.

By considering a larger number of couplings $t_\mathbf{m}$, we can include long-range couplings to the 2D-SSH model.
We observe the topological states ($\delta > 0, \Delta m = 0$) to be robust, which is evident when considering that no new band touchings were generated as the distance of couplings $M$ is increased.
As chiral symmetry is broken, the energies of the edge modes are shifted toward the bulk states, see Methods.

A vectorized 2D-SSH model can be obtained by using the expanded $\hat H_k$ from Eq. \ref{eq:Hk_vec} and $\hat{H}_c$ built from vectorized representation matrices as shown by Eq. \eqref{eq:proj_vec}.
The dispersion relation of a vectorized model with $M = 1$ is shown in Fig. \ref{fig:vectorization} (b).
Due to the doubling of degrees of freedom, the number of modes has also doubled.
The symmetry properties of the vector modes are the same as in \cite{kim2020topological}; however, our dispersions are slightly different due to additional couplings.

\subsection*{Connecting local and global topology}
The conventional and radiation Berry curvatures are not strictly equivalent due to the presence of far-field polarization singularities~\cite{yin2025observation, yuan2025breakdown}. 
These singularities are connected to a notion of local topology, where topological defects manifest as disclinations or vortices in the phase or the polarization for scalar and vector fields, respectively.
Vortices of polarization (in momentum space) are associated with BICs, characterized by a winding number $q$ defined on a circular path $C$ as
\begin{equation*}
    q = \frac{1}{2\pi}\oint_C d\mathbf{k}\cdot\nabla \phi(\mathbf{k}),
\end{equation*}
where the polarization angle $\phi(\mathbf{k}) = \frac{1}{2}\arg(S_1(\mathbf{k}) + iS_2(\mathbf{k}))$ is defined from the Stokes parameters corresponding to linear polarization. 
The winding number $q$ is also known as the index of disclination in the literature on structured light beams.
The topological charge is connected to the phase singularities in the left- and right-handed components of the field $q = \frac{q_- - q_+}{2}$~\cite{yoda2020generation}, where 
\begin{equation*}
    q_\pm = \frac{1}{2\pi}\oint_C d\mathbf{k}\cdot \nabla \arg(E_x(\mathbf{k}) \pm iE_y(\mathbf{k})).
\end{equation*}

At HSPs, the polarization vortices are symmetry-protected and belong to different IRs based on their rotational properties.
The construction of $\hat{H}_c$ in our method allows us to arbitrarily reorder these IRs to different bands.
This creates bands that contain combinations of $q_\pm$ at the various HSPs, allowing us to study the connection between topological defects and global topological invariants.

\begin{figure*}
    \centering
    \includegraphics[width=0.95\textwidth]{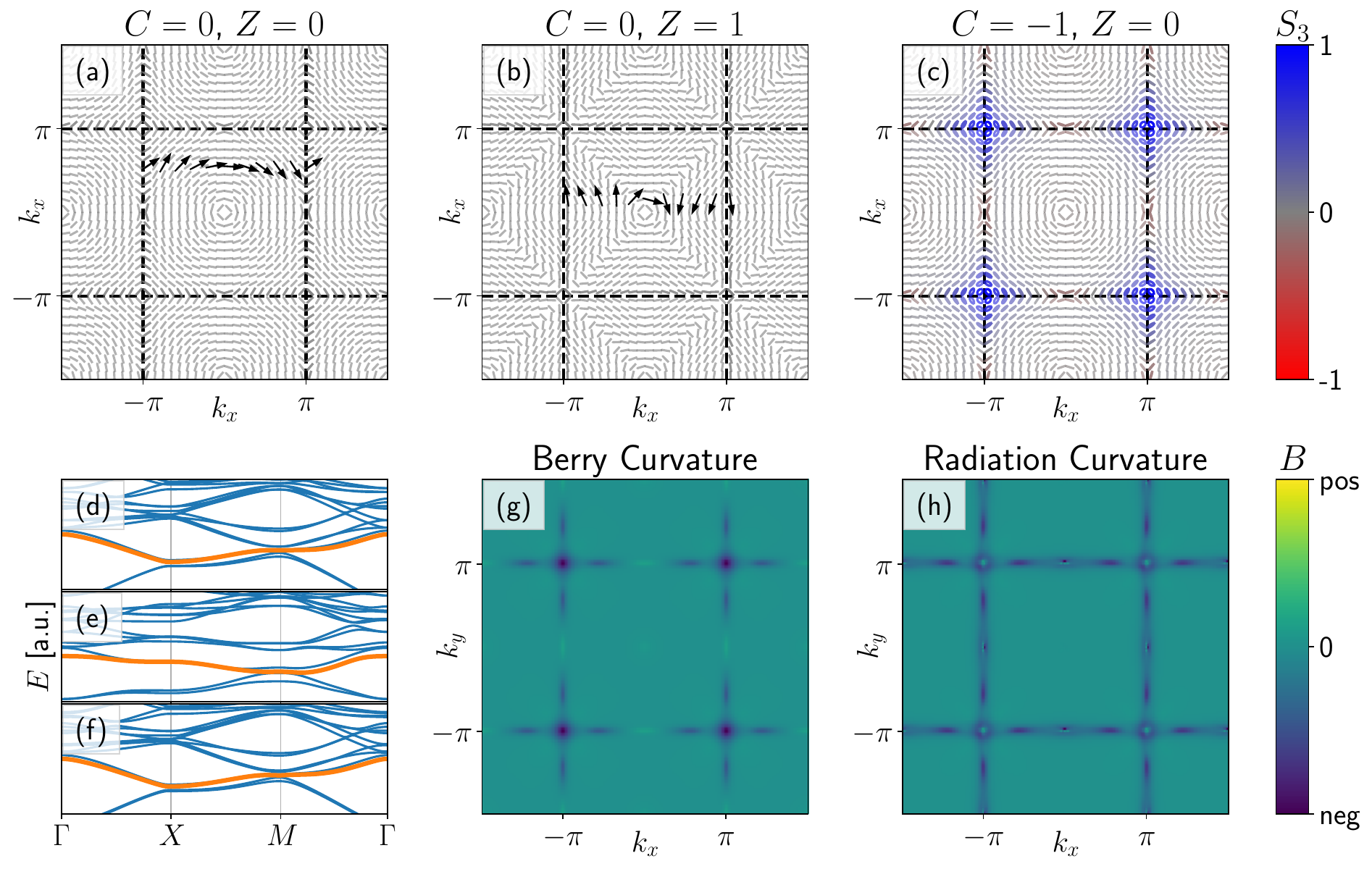}
    \caption{
        (a-c) Far field polarization patterns for three different configurations of $\hat{H}_c$, resulting in the third band having three distinct sets of topological invariants.
        All of the patterns have the same vortex in the center of the BZ.
        The topological invariants are calculated directly from $\ket{\psi(\mathbf{k})}$.
        (d-f) The corresponding band structures with the band of interest highlighted in orange.
        Note that in (d,f), there is a second band very close to the orange one.
        In (a/d), the band is topologically trivial ($C=0)$ without any band polarization ($Z=0$).
        This band has a band-touching point at $M$, but it remains topologically trivial even when the upper band is considered.
        In (b/e), the band has a net band-polarization ($Z=1$), as is visible from having a vortex at $\Gamma$ and a trivial polarization at $X$ and $Y$.
        The arrows in panels (d,e) show the winding of the electric field on a closed loop traversing the BZ, highlighting the effect of a nontrivial Zak phase.
        The band in (c/f) carries a Chern number related to the open gap at $M$.
        This is seen as the polarization picking up a circular component, leaving the band otherwise unaffected, as compared to (a/c).
        The topological charges of the polarization vortices cancel each other in all cases, leading to a net zero topological charge.
        (g,h) The Berry and radiation curvatures are calculated for the case shown in panels (c/f).
        The full energies and eigenvectors are accessible at \cite{full_data}.
    }
    \label{fig:ff_comparison}
\end{figure*}

We design real-space couplings that realize combinations of polarization vortices leading to either topologically trivial or non-trivial bands.
As an example, we consider the square lattice with vector fields and various coupling Hamiltonians $\hat{H}_c$. 
Specifically, we choose the number of sites in the unit cell to be $16$ ($N_x=N_y=4$) and include $M = 8$ Fourier components in the dispersion. 
We study the topology of the third lowest band, highlighted in orange in Fig.\ref{fig:ff_comparison}(d-f).

The first configuration shown in Fig.~\ref{fig:ff_comparison}(a,d) considers a topologically trivial band; the parameters are $\Delta\varepsilon_{A_1} = -0.5, \Delta\varepsilon_{A2} = -0.8, \Delta\varepsilon_{B1} = 1.3, \Delta\varepsilon_{B2} = 1.2, \Delta\varepsilon_E = 0.1$.
Due to the increased number of bands, the crossings at $\Gamma$ and $M$ now contain two $E$-doublets which cannot be separated with $\hat{H}_c$ alone.
To lift the degeneracy between the two $E$ doublets, we include an additional term $\hat H_\parallel$, as outlined in the Methods, with $\Delta\varepsilon_\parallel = -0.15$.
In essence, this term couples the parallel field components between adjacent dipoles.
At the $M$-point, the trivial configuration has an $E$-doublet, which can be seen as a degeneracy in the dispersion.
The far-field pattern displays topological defects, but all states have pure linear polarization, as is shown in Fig.~\ref{fig:ff_comparison}(a).
When calculating the various charges at the HSPs, we find that $q_\Gamma = q_M = 1$ and $q_X = q_Y = -1$; hence, the sum of all charges is zero.
Despite the presence of nonzero local topological charges, the band topology is trivial as both the Zak phase $Z$ and the Chern number $C$ are zero. 
The absence of the Zak phase can be seen by tracking the evolution of an eigenstate $\ket{E(\mathbf{k})}$ on a path crossing the BZ.
This can be seen, for instance, by tracking the evolution of the eigenstate $\ket{E(\mathbf{k})}$ on a path crossing the BZ, highlighted by black arrows in Fig.~\ref{fig:ff_comparison}(a).
As the state returns to its original orientation, there is no net phase accumulation.
Additionally, as the system has both TRS and inversion symmetry, and we are ignoring non-Hermitian effects, the Berry curvature (and thus the Chern number) is identically zero~\cite{gong2018topological,kawabata2019symmetry,hu2024generalized, cuerda2024observation, cuerda_pseudospin-orbit_2024, yuan2025breakdown}.

A different situation is analyzed in Fig.~\ref{fig:ff_comparison} (b,e), for $\Delta\varepsilon_{A_1} = -2.4, \Delta\varepsilon_{A2} = -3, \Delta\varepsilon_{B1} = -2.3, \Delta\varepsilon_{B2} = -2.0, \Delta\varepsilon_E = -1$ with additional parallel coupling of $\Delta\varepsilon_\parallel = -0.5$ for dipoles.
Focusing again on the third band, the far-field states are still all linearly polarized. 
However, in contrast with the previous case, there is now a mismatch in the inversion symmetry-eigenvalues at $\Gamma$ and $X$, leading to a nontrivial Zak phase.
Due to the $\mathbf{C}_4$ symmetry, the Zak phases evaluated along $x$ and $y$ directions are the same $Z_x = Z_y \equiv Z$.
By tracking the orientation of the electric field over a cut across the BZ in Fig.~\ref{fig:ff_comparison}(b), we see a $\pi$ phase winding accumulation, contrasting with the trivial ($Z=0$) configuration discussed above.
The topological charge $q_\Gamma=1$ is the same as before, but the symmetry change at $X$ causes the polarization vortex to disappear, $q_X=0$.
The topological charge $q_M=-1$ ensures the total charge is zero. 
As before, Berry curvature is identically zero.

We can now highlight a connection between the topological charges $q$ and the Zak phase.
The Zak phase is uniquely defined by the behavior of modes at $\Gamma$ and $X$ under inversion symmetry ($\mathbf{C}_2$ in 2D) \cite{fang2012bulk,vaidya2023topological}.
In terms of topological charges, modes that are even under inversion have odd charges and vice versa.
Hence, a symmetry-mismatch occurs when one of the charges is even and the other is odd, i.e., $Z = (q_\Gamma + q_X)\mod 2$.
Note that as the charges sum up to zero, $q_\Gamma$ and $q_X$ uniquely define $q_M$.

The third configuration in Fig.~\ref{fig:ff_comparison}(c,f) considers the same parameters as in Fig.~\ref{fig:ff_comparison}(a,d), but with the inclusion of the TRS breaking term $\hat H_m = \Delta m (P_{E_1} - P_{E_2})$, with $\Delta_m = 0.05$.
This has the effect of splitting the band degeneracy at $M$ into circularly polarized states, see Fig.~\ref{fig:ff_comparison}(f), where the blue (red) color indicates left (right) circular polarization.
The Berry curvature in the near-field $B_\text{NF}(\mathbf{k})$ is mainly concentrated around the $M$-point, where the bands are closest to one another. This is consistent with the fact that the near-field Berry curvature is most sensitive to regions of band proximity and hybridization between eigenstates.
In contrast, the Berry curvature from the far field shows a different distribution. 
This is because, in addition to band proximity, as indicated by Eq.~\eqref{eq:berryff}, the far-field Berry curvature reflects the variation in the degree of circular polarization, which remains nearly constant around the $M$-point and thus leads to small far-field Berry curvature. 
On the other hand, towards the $X$ and $Y$ points, the polarization becomes increasingly linear, decreasing the Berry curvature that is associated with a circular component. 
As a result, the far-field Berry curvature is seen to peak in between the $M$ and $X$/$Y$ points.
The topological charge $q = + 1$ at $M$ remains unchanged, but left circular emission indicates that one of the polarizations is trivial, i.e., $q_- = 0$ and hence $q_+ = -2$.
While the distribution of the conventional and radiation curvatures differs, both yield the same Chern number, confirming that while the local topology depends on whether observed in the far or near field, the global topology is robustly encoded in both the near and far field.

\begin{figure*}
    \centering
    \includegraphics[width=\linewidth]{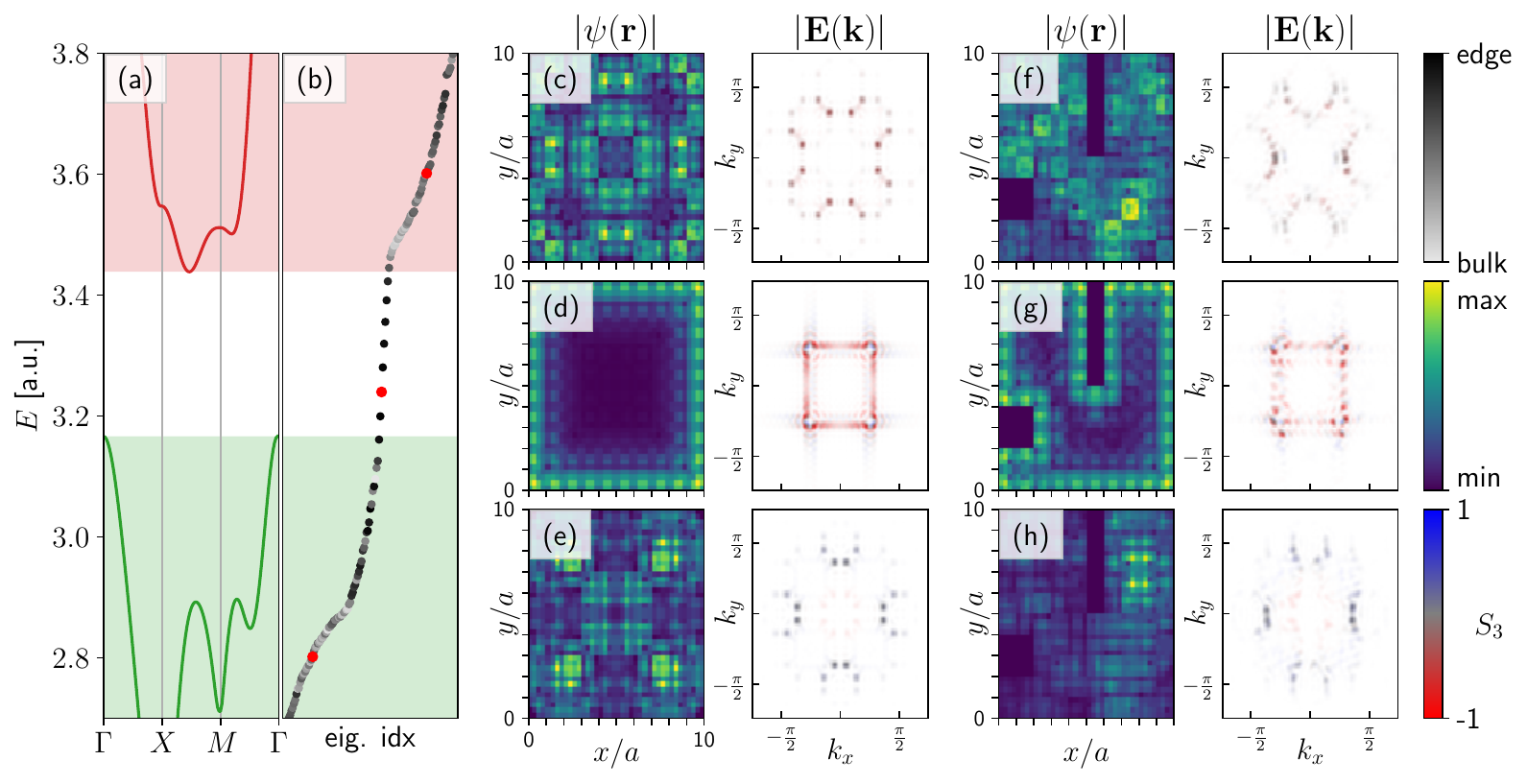}
    \caption{
        (a) A close-up of the dispersions of the bulk bands in the region of interest, displaying a clear, direct band gap.
        The bands shown are topologically equivalent to the systems discussed in Figs. \eqref{fig:ff_comparison} (c/f) and \eqref{fig:ff_comparison2} with the couplings exaggerated even further.
        The shaded regions display the range of energies covered by the bulk bands.
        (b) Sorted eigenvalues of a Hamiltonian with open boundary conditions (10x10 unit cells, with $N = 4$, $M = 8$) in the same energy range.
        A set of edge modes (grey-black denote bulk and edge modes, respectively) can be seen crossing the gap.
        Modes shown at (c-h) are highlighted with red dots.
        (c-e) The real-space distributions of eigenmodes highlighted in (b) with their corresponding far fields.
        The far-field is colored based on the $S_3$ component of the polarization.
        The bulk modes (c,e) show weak dots with small circular polarization.
        The topological edge mode (d) is localized around the sample edge, with most of the mode existing in the outermost unit cells.
        The far-field emission can be seen to be fully circularly polarized, and the emission draws a closed contour in $k$-space as opposed to the speckles seen from the bulk modes.
        (f-h) Modes at similar energies to (c-e) but from a system where a portion of the sample is removed.
        The topological edge mode is seen to follow the edge of the sample despite any irregularities in the shape.
        The far field emissions can be seen to spread out and break up a bit, but the emission from the topological edge state stays mostly the same.
        The full energies and eigenvectors are accessible at \cite{full_data}.
    }
    \label{fig:finite_emissions}
\end{figure*}

As a final test of the model, we generate a system with open boundary conditions in $x$ and $y$.
The bulk bands of interest are topologically equivalent to the structure in Fig.~\ref{fig:ff_comparison} (c/f) and with the couplings exaggerated to make the band-gap direct.
By solving the eigenvalues, we can identify the edge-states crossing the band-gap, as is shown in Fig.~\ref{fig:finite_emissions}.
We can now look for far-field signals that indicate that the emission is from a topological edge state.
By comparing the emission to two representative bulk modes, we can see that the far field of a topological mode is fully spin-polarized.
In addition, the emission draws full contours in momentum space that match the edge shape of the sample, even when sharp features (defects) are present at the edge, demonstrating the topological protection of the edge state.

\section*{Discussion}
We have developed a real-space effective Hamiltonian that applies to both scalar and vector fields and is continuously deformable from the TB limit, with nearest-neighbor couplings, to the photonic ELA limit, with long-ranged couplings.

We applied the effective Hamiltonian to describe a family of 2D SSH-like models. We observed that the topological quantities are invariant when the range of couplings is increased, as no new crossings are generated. Our approach allows us to understand the connection between local topological defects, characterized by polarization singularities, and the global topological invariants of the photonic bands. 
By calculating the far fields, we found signals from the HSPs of the BZ that indicate the presence of global topological invariants. 
Both in the far- and near-field cases, the Zak phase appears when the symmetry properties of the modes at $\Gamma$ and $X$ are mismatched.
We showed that in terms of topological charges $q$ this condition is $Z/\pi = q_\Gamma + q_X \mod 2$ for $\mathbf{C}_4$. 
This result is generalized to $\mathbf{C}_2$ as $Z_{x,y}/\pi = q_\Gamma + q_{X,Y} \mod 2$ as the charges at $X$ and $Y$ are allowed to be different.
The band obtains a Chern number when a degeneracy point is split by breaking TRS.
In the far field, this appears as circularly polarized states.
We observed that the topological charge $q$ remains invariant when the values of the global topological invariants $C$ and $Z$ change, indicating that the Chern number is connected, instead, to the imbalance between the handedness-weighed sub-charges $q_\pm$.
Interestingly, the near- and far-field Berry curvature patterns are different, despite yielding the same Chern number. Although we focused on the Berry curvature here, our approach readily enables the study of the full quantum geometry of the system, including the quantum metric~\cite{resta2011insulating,Torma2023}. 

Because our effective Hamiltonian is constructed analytically from symmetry considerations, it provides access to real-space couplings. The real-space Hamiltonian can be recovered via inverse Fourier transform, which for the constant coupling Hamiltonian $\hat H_c$ is trivial, and directly implemented in a variety of experimental platforms. 
In photonic systems, it gives a good guideline for the design of the unit cell and lattice geometry. However, values of the generated real-space hoppings may need to be fine-tuned by numerical simulations. 
In synthetic matter platforms, the real space Hamiltonian can be directly realized: promising setups include circuit lattices, mechanical metamaterials, and reconfigurable cavity arrays, where each site is a discrete resonator and couplings are implemented via passive elements such as lumped inductors and capacitors, or springs. 
Our effective Hamiltonian approach, therefore, provides an easy-to-use framework for designing topologically non-trivial lattices across a wide range of photonic and synthetic platforms.
Alternatively, the approach can also serve as a symmetry-guided TB effective Hamiltonian of more complicated systems, where the fitting parameters, namely the deviation of IRs from the empty-lattice energy at HSPs, are easily obtainable from more computationally demanding simulations or from experiments.
In this case, its role is to provide physical understanding, intuition, and a connection to known solid-state models and phenomena. 
A third way to apply the effective Hamiltonian is to use it as a simple but powerful theoretical framework for understanding connections between parameter regimes and phenomenology that are usually described by distinct theoretical tools -- the connections between local and global topology studied here are an example of this.

\section*{Methods}
\subsection*{2D momentum-dependent Hamiltonian}
In 2D, we can express $\hat{H}_k$ in terms of hoppings between sites inside a single unit cell as
\begin{equation}
    \hat{H}_k = \sum_{i,j,i',j'} H_{ij,i'j'}\hat{a}_{i,j}^\dagger \hat{a}_{i',j'},
\end{equation}
where $i,i'$ and $j,j'$ correspond to the $x$- and $y$ indices.
In general, the matrix elements $H_{ij,i'j'}$ form as a sum
\begin{align}
    &H_{ij,i'j'} = 
    \sum_{\mathbf{m}}t_{\mathbf{m}} e^{i(\Delta_x k_x + \Delta_yk_y)}\delta_{(\Delta_i - m_x)\text{ mod }  N} \delta_{(\Delta_j - m_y)\text{ mod }N}
    \label{eq:2d_lrEla}
\end{align}
with $\Delta_{x} = \lfloor (m_x + i')/N\rfloor$ and$ \Delta_y = \lfloor (m_y + j')/N\rfloor$ to the distance between unit cells with $N$ being the length of a side of the unit-cell and $m_x,m_y$ steps along their respective directions; and $\Delta_{i} = i-i'$, $\Delta_{j} = j-j'$ denotes over how many sites in the unit cell the hopping occurs. 
As the couplings are only determined by the distance, all $t_{\mathbf{m}}$ with $m_x^2 + m_y^2 = |\mathbf{m}|^2$ are equal.

\subsection*{Symmetry groups}
The $\mathbf{C}_n$ group is generated by a $2\pi/n$-degree rotation and has $n$ elements.
The properties of the symmetry groups are collected in character tables, where the characters show the behavior of irreducible representations (IRs) under different symmetry operations.
Symmetry elements are physically understood as operators, while IRs and characters are their eigenvectors and eigenvalues, respectively.
For information on how these tables are built for other symmetries, we refer to relevant literature~\cite{inui2012group}.

The character table provides two powerful concepts: the IR decomposition and the IR projection matrices.
The IR decomposition allows us to identify the number of modes belonging to each IR in the system without ever touching a Hamiltonian.
This gives information on the rotational characteristics of the modes and the number of symmetry-protected degeneracies.
The number of IRs is found by calculating 
\begin{equation}
    n_\text{IR} = \frac{1}{g}\sum_i \epsilon_{i,IR}\text{Tr}(D(g_i)),
    \label{eq:irrep_decomp}
\end{equation}
where $i$ sums over all symmetry elements $g_i$ and $g$ is the size of our group.
Here $D(g_i)$ is the representation of $g_i$ in our basis, i.e., a square matrix that does the operation $g_i$, and $\epsilon_{i,\text{IR}}$ are the elements of the character table.

\subsection*{C${}_4$ symmetry}
Here we provide explicit formulas for performing IR decomposition and building projection matrices for a $\mathbf{C}_4$ symmetric system. 

The $\mathbf{C}_4$ group is generated by a $\pi/2$-degree rotation and consists of four elements ($\{E,C_4,C_2,C_4^3\}$) corresponding to rotations by $0,\pi/2,\pi$ and $3\pi/2$ clockwise respectively.
The properties of the $\mathbf{C}_4$ symmetry group are collected in Table~\ref{tab:char_c4}.

\begin{table}[h]
    \centering
    \begin{tabular}{c|c|c|c|c}
        $\mathbf{C}_4$ &  $E$ & $C_4$ & $C_2$ & $C_4^3$ \\
        \hline
        $A$ & $1$ & $1$ & $1$ & $1$ \\
        \hline
        $B$ & $1$ & $-1$ & $1$ & $-1$ \\
        \hline
        $E_1$ & $1$ & $i$ & $-1$ & $-i$ \\
        $E_2$ & $1$ & $-i$ & $-1$ & $i$ \\
    \end{tabular}
    \caption{Character table of the $\mathbf{C}_4$ group.}
    \label{tab:char_c4}
\end{table}   

Let us consider the operators that perform the clockwise rotations of the sites in a unit cell of a scalar 2D SSH model, as in the main text:

\begin{equation}
\begin{split}
    &D(E) = \begin{bmatrix}
1 & 0 & 0 & 0\\
0 & 1 & 0 & 0\\
0 & 0 & 1 & 0\\
0 & 0 & 0 & 1\
\end{bmatrix}, \quad
 D(C_4) = \begin{bmatrix}
0 & 1 & 0 & 0\\
0 & 0 & 0 & 1\\
1 & 0 & 0 & 0\\
0 & 0 & 1 & 0\\
\end{bmatrix},\\
&D(C_2) = \begin{bmatrix}
0 & 0 & 0 & 1\\
0 & 0 & 1 & 0\\
0 & 1 & 0 & 0\\
1 & 0 & 0 & 0
\end{bmatrix} \quad
D(C_4^3) =  \begin{bmatrix}
0 & 0 & 1 & 0\\
1 & 0 & 0 & 0\\
0 & 0 & 0 & 1\\
0 & 1 & 0 & 0
\end{bmatrix}.
\end{split}
\end{equation}

The projection matrices are calculated with Eq.~\eqref{eq:projections} in the main text
\begin{equation}
\begin{split}
    &P_A = \frac{1}{4}\begin{bmatrix}
1 & 1 & 1 & 1\\
1 & 1 & 1& 1\\
1 & 1 & 1 & 1\\
1 & 1 & 1 & 1
\end{bmatrix}, \; P_B = \frac{1}{4}\begin{bmatrix}
1 & -1 & -1 & 1\\
-1 & 1 & 1& -1\\
-1 & 1 & 1 & -1\\
1 & -1 & -1 & 1
\end{bmatrix}, \\
&
P_{E_1} = \frac{1}{4}\begin{bmatrix}
1 & i & -i & -1\\
-i & 1 & -1& i\\
i & -1 & 1 & -i\\
-1 & -i & i & 1
\end{bmatrix},\; P_{E_2} = \frac{1}{4}\begin{bmatrix}
1 & -i & i & -1\\
i & 1 & -1& -i\\
-i & -1 & 1 & i\\
-1 & i & -i & 1  
\end{bmatrix} .
\end{split}
\label{eq:projection_matrices_c4}
\end{equation}
The projection matrices in Eq.~\eqref{eq:projection_matrices_c4} are to be used for constructing the coupling Hamiltonian in Eq.~\eqref{eq:H_c} in the main text.

\subsection*{C${}_{4v}$ symmetry}

Here we provide explicit formulas for performing IR decomposition and building projection matrices for a $\mathbf{C}_{4v}$ symmetric system, used for the 2D SSH system of vector fields. 
The character table for $\mathbf{C}_{4v}$ is reported in Table~\ref{tab:char_c4v}.

\begin{table}
\begin{tabular}{c|c|c|c|c|c|c|c|c}
        $\mathbf{C}_{4v}$ &  $E$ & $C_4$ & $C_2$ & $C_4^3$ & $\sigma_x$ & $\sigma_y$ & $\sigma_d$ & $\sigma_d'$\\
        \hline
        $A_1$ & $1$ & $1$ & $1$ & $1$ & $1$ & $1$ & $1$ & $1$ \\
        \hline
        $A_2$ & $1$ & $1$ & $1$ & $1$ & $-1$ & $-1$ & $-1$ & $-1$ \\
        \hline
        $B_1$ & $1$ & $-1$ & $1$ & $-1$ & $1$ & $1$ & $-1$ & $-1$ \\
        \hline
        $B_2$ & $1$ & $-1$ & $1$ & $-1$ & $-1$ & $-1$ & $1$ & $1$ \\
        \hline
        $E$ & $2$ & $0$ & $-2$ & $0$ & $0$ & $0$ & $0$ & $0$ \\
    \end{tabular}
    \caption{Character table of the $\mathbf{C}_{4v}$ group.}
    \label{tab:char_c4v}
\end{table}  

In addition to rotations we have reflections $\sigma_x,\sigma_y,\sigma_d$ and $\sigma_d'$ which reflect over $y$ and $x$ axes and diagonals.
On the chosen basis, they have the representations
\begin{equation}
\begin{split}
    D(\sigma_x) = \begin{bmatrix}
0 & 0 & 1 & 0\\
0 & 0 & 0 & 1\\
1 & 0 & 0 & 0\\
0 & 1 & 0 & 0\\
\end{bmatrix}, D(\sigma_y) = \begin{bmatrix}
0 & 1 & 0 & 0\\
1 & 0 & 0 & 0\\
0 & 0 & 0 & 1\\
0 & 0 & 1 & 0\\
\end{bmatrix}, \\ D(\sigma_d) = \begin{bmatrix}
1 & 0 & 0 & 0\\
0 & 0 & 1 & 0\\
0 & 1 & 0 & 0\\
0 & 0 & 0 & 1\\
\end{bmatrix}D(\sigma_d') = \begin{bmatrix}
0 & 0 & 0 & 1\\
0 & 1& 0 & 0\\
0 & 0 & 1 & 0\\
1 & 0 & 0 & 0\\
\end{bmatrix}.
\end{split}
\end{equation}
Calculating traces for IR decomposition we find that $\text{Tr}(D(\sigma_x)) = \text{Tr}(D(\sigma_y)) = 0, \text{Tr}(\sigma_d) = \text{Tr}(\sigma_d') = 2$.
Using equation \ref{eq:irrep_decomp} we find
\begin{equation}
\begin{split}
    n_{A_1} = \frac{1}{8}\cdot8 = 1, \hspace{0.2cm}n_{A_2} = \frac{1}{8}\cdot 0 = 0, \hspace{0.2cm}
    n_{B_1} = \frac{1}{8}\cdot 0 = 0,\hspace{0.2cm}  \\ n_{B_2} = \frac{1}{8}\cdot 8 = 1,\hspace{0.2cm}
    n_{E} = \frac{1}{8}\cdot 8 = 1. 
    \end{split}
\end{equation}
The system contains the singlets $A_1$ and $B_2$ and a doublet $E$, remaining consistent with the group $\mathbf{C}_4$.
The presence of reflectional symmetries makes $E_1$ and $E_2$ equivalent, as now there is no difference between left and right rotation.

The new projection matrices for the IRs are:
\begin{equation}
\begin{split}
    P_{A_1} = \frac{1}{8} \begin{bmatrix}
2 & 2 & 2 & 2\\
2 & 2 & 2& 2\\
2 & 2 & 2 & 2\\
2 & 2 & 2 & 2\\
\end{bmatrix}, \hspace{0.4cm}
 P_{B_2} = \frac{1}{8} \begin{bmatrix}
2 & -2 & -2 & 2\\
-2 & 2 & 2& -2\\
-2 & 2 & 2 & -2\\
2 & -2 & -2 & 2\\
\end{bmatrix} , \\
P_E = \frac{2}{8} \begin{bmatrix}
2 & 0 & 0 & -2\\
0 & 2 & -2& 0\\
0 & -2 & 2 & 0\\
-2 & 0 & 0 & 2\\
\end{bmatrix}.
\end{split}
\label{eq:proj_matrices_c4v}
\end{equation}

\subsection*{Connection to 2D Su-Schrieffer-Heeger model}

We shall now apply our framework to the 2D Su-Schrieffer-Heeger model in the tight-binding approximation (2D-SSH). The momentum-dependent part of the Hamiltonian matrix can be obtained with $N_x=N_y = 2$ and $t_\mathbf{m} = t\cdot\delta_{|\mathbf{m}|^2, 1}$ as
\begin{equation}
    \hat{H}_k= t \begin{bmatrix}
        0 & 1+ e^{i k_y} & 1+ e^{i k_x} & 0 \\
        1+ e^{-i k_y} & 0 & 0 & 1+ e^{-i k_x}\\
        1+ e^{-i k_x} & 0 & 0 & 1+e^{-i k_y} \\
        0 & 1+e^{i k_x} & 1+ e^{i k_y} & 0       
    \end{bmatrix}
    \label{eq:2dsshtbEla}
\end{equation}
Notice that the inter- and intra-cell couplings $t$ are the same, since $\hat H_k$ gives the tight-binding dispersion with closed gaps. 
The band structure of Eq.~\eqref{eq:2dsshtbEla} is plotted in Fig.~\ref{fig:model_evolution}(b).

This model is $\mathbf{C}_{4v}$ symmetric, meaning that the unit cell is invariant under rotations and reflections.
However, the model can be fully constructed from $\mathbf{C}_4$, implying that the reflectional symmetries do not play a role.
In addition, working with $\mathbf{C}_4$ allows for the inclusion of time-reversal-symmetry (TRS) breaking (magnetic) effects.
Using Eq.~\eqref{eq:H_c} together with the projection matrices for $\mathbf{C}_{4}$ symmetry, we calculate the coupling Hamiltonian $\hat{H}_c$:
\begin{equation}
\hat{H}_{c} = 
    \begin{bmatrix}w_+ & v & v^{*} & w_-\\
    v^{*} & w_+ &  w_- & v\\
    v & w_- &  w_+ & v^{*}\\
    w_- & v^{*} & v & w_+
    \label{eq:2dsshcoupling}
    \end{bmatrix},
\end{equation}
with $v= \Delta \varepsilon_{A} - \Delta \varepsilon_{B} + i \left(\Delta \varepsilon_{E_1}- \Delta \varepsilon_{E_2}\right)$ and $w_\pm = \Delta \varepsilon_{A} + \Delta \varepsilon_{B} \pm (\Delta \varepsilon_{E_1} + \Delta \varepsilon_{E_2})$. 
Note that if $\mathbf{C}_{4v}$ was used instead, $\hat H_c$ would be real as the projection matrices in Eq. \eqref{eq:proj_matrices_c4v} are real.
By separately tuning the IRs energies $\Delta \varepsilon_{E_1}, \Delta \varepsilon_{E_2}, \Delta \varepsilon_A, \Delta \varepsilon_B$, different topological regimes can be obtained. 
In particular, if $\Delta \varepsilon_{E_1} =\Delta \varepsilon_{E_2} =0$ and $\Delta \varepsilon_A= -\Delta \varepsilon_B = \delta/2$, the usual 2D-SSH Hamiltonian~\cite{liu2017novel} is recovered
\begin{equation}
\hat{H}(k_x, k_y)= \hat{H}_k + \hat{H}_c = \begin{bmatrix}
        0 & t_1+ t e^{i k_y} & t_1+ t e^{i k_x} & 0 \\
        t_1+ te^{-i k_y} & 0 & 0 & t_1+ te^{-i k_x}\\
        t_1+ te^{-i k_x} & 0 & 0 & t_1+te^{-i k_y} \\
        0 & t_1+te^{i k_x} & t_1+ te^{i k_y} & 0       
    \end{bmatrix}
    \label{eq:2dsshtotal}
\end{equation}
where $t_1 = t+\delta$, so that the topological transition occurrs at $\delta = 0$. Note that from the inverse Fourier transform, $t < 0$.

\begin{figure}
    \centering
    \includegraphics[width=0.5\linewidth]{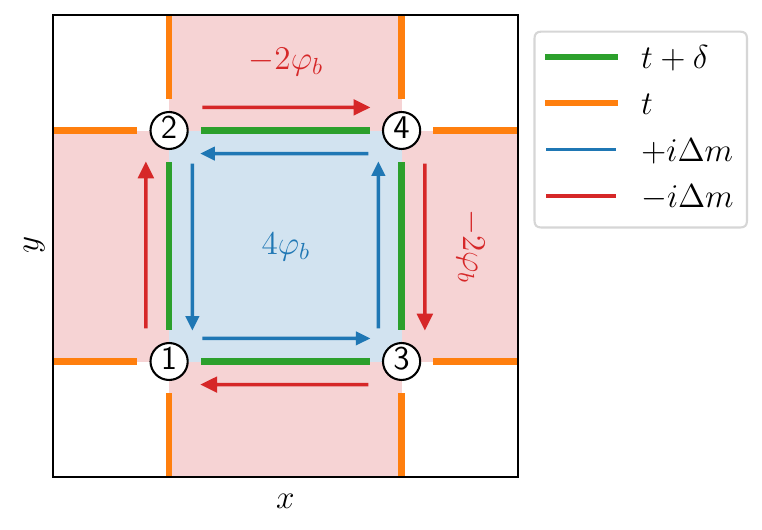}
    \caption{The 2D SSH model unit cell with all relevant couplings in the TRS-breaking (magnetic) case.}
    \label{fig:fun_ssh_uc}
\end{figure}

When $\Delta \varepsilon_{E_1} =- \Delta \varepsilon_{E_2}$, and $\Delta \varepsilon_{E_1} - \Delta \varepsilon_{E_2} = \Delta_m $, the intracell couplings in Eq.~\eqref{eq:2dsshtotal} become complex $t_1 = |t'| e^{i \varphi_m}$, with a Peierls phase $\varphi_m =  \arctan\left[\Delta_m/(t+\delta)\right]$, and $|t'| = \sqrt{(t+\delta)^2+\Delta_m^2}$, while the intercell couplings $t$ remain real. Such a situation breaks TRS, and corresponds to a flux lattice, where alternating plaquettes have magnetic fluxes $4\varphi_m$ and $-2\varphi_m$ respectively, see Fig.~\ref{fig:fun_ssh_uc}.

\begin{figure*}
    \centering
    \includegraphics[width=0.7\textwidth]{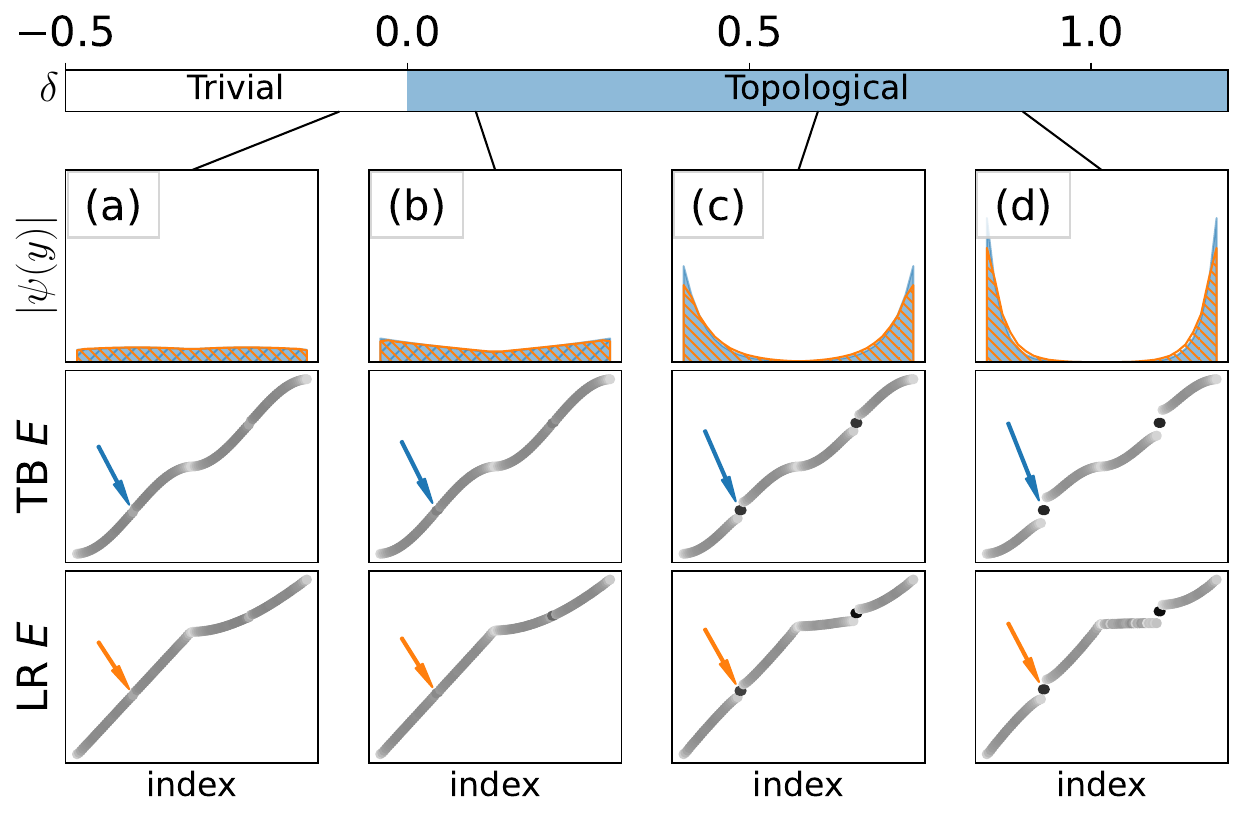}
    \caption{
    Topology of the 2D SSH model with open boundary conditions in the y-direction, both in the tight-binding (TB) and the long-ranged (LR) couplings case when $\delta$ is varied with (a) being trivial and (b-d) topological, with $\Delta_m=0$. 
    Since the inclusion of longer-range couplings only distorts the ELA dispersion but does not open new gaps, the topological transition is unmodified, and edge states are found both for the tight-binding and the photonic-like dispersion. 
    The top row shows the state wavefunction in the tight-binding (blue) and the long-range (orange) case, with the modes of interest highlighted in their respective eigenvalue spectra, where the color indicates the edge-bulk nature of the mode (dark-light, respectively).
    }
    \label{fig:2dsshtopo}
\end{figure*}

\subsubsection*{Long-ranged}
By including the Fourier coefficients $t_\mathbf{m}$ up to $ M$th order (with $N_x=N_y = 2$), we describe the long-ranged regime, and smoothly approach the exact ELA dispersion at $M = \infty$.
This evolution is shown in Fig.~\ref{fig:model_evolution}(b$\rightarrow$e$\rightarrow$h).

The band topology is determined by the structure of the coupling Hamiltonian $\hat{H}_c$, which is the only term that can form new symmetry-induced gap openings.
We find that the topological transition of the full model emerges regardless of whether the underlying dispersion is tight-binding or long-range, see Fig.~\ref{fig:2dsshtopo}. 
The top row in Fig.~\ref{fig:2dsshtopo} shows the formation of the edge state from the topologically trivial ($\delta<0$ in Fig.~\ref{fig:2dsshtopo}(a)) to the nontrivial ($\delta>0$ in Fig.~\ref{fig:2dsshtopo}(b-d)) region, for both the tight-binding (blue) and the long-ranged (orange) case. 
The second and third row in Fig.~\ref{fig:2dsshtopo} show the energy spectra, ordered by the eigenvalue number of the Hamiltonian with open boundary conditions, where the colorscale indicates how much the state is localized near the edge.
From the spectra, we notice that while the tight-binding model exhibits chiral symmetry, long-range couplings break this symmetry and edge states are no longer pinned in the middle of the band gap but rather shift towards the bulk states, see Fig.~\ref{fig:2dsshtopo}.

\subsubsection*{Vector fields}
We now consider the vector extension of the 2D SSH model
When vectorizing the model, it is necessary to use the representation matrices of the $\mathbf{C}_{4v}$-group, as the reflection symmetries provide couplings between the $L$ and $R$ components, see Fig.~\ref{fig:vectorization}.
This is evident when looking at Eq.~\eqref{eq:repr_vec}, where the representation matrices $D_v(g_i)$ are block-diagonal for rotations due to $D'(g_i)$ being diagonal, see Methods
The off-diagonal elements only come from the representation matrices for reflections, see Methods.
TRS-breaking (magnetic) effects can be included in vector lattices by using the $\mathbf{C}_4$ and $\mathbf{C}_{4v}$ symmetries together.
As was the case with scalar fields, we can construct a TRS-breaking coupling Hamiltonian in $\mathbf{C}_4$
\begin{equation}
    \hat{H}_m = \Delta_m\left(P_{E_1} - P_{E_2} \right),
\end{equation}
with $\Delta \varepsilon_{E_2} = -\Delta \varepsilon_{E_1} \equiv \Delta_m/2$, which makes $H_m$ purely imaginary.
Now, the total coupling Hamiltonian for the vector case with broken TRS is
\begin{equation}
    \hat{H}_c =\hat{H}_m + \sum_{i\in \mathbf{C}_{4v}} \Delta \varepsilon_i P_i.
\end{equation}

A comparison between the scalar and vector model in a configuration similar to the 2D SSH model with long-range couplings is shown in Figure \ref{fig:vectorization}. The vector model has twice the number of bands of the scalar model, since the polarization degree of freedom is now included. 
Examples of IRs are also shown, where the polarization on each site is depicted as a vector.

\subsection*{Representation matrices of in-plane vectors}
Working in the circularly polarized basis, the representation matrices for rotations of $\alpha$ are
\begin{align}
    D'(g_\alpha) = \begin{bmatrix}
        e^{i\alpha} & 0\\ 0 & e^{-i\alpha}
    \end{bmatrix}.
    \label{eq:rot_vec}
\end{align}
Similarly, the representation matrix for reflections over the axis at angle $\theta$ is

\begin{align}
    D'(g_\theta) = \begin{bmatrix}
        0 & e^{-i2\theta}\\ e^{i2\theta} & 0
    \end{bmatrix} .
    \label{eq:refl_vec}
\end{align}

\subsection*{Parallel coupling Hamiltonian $\hat H_\parallel$}
\label{sec:H_par}

As mentioned in the main text, a constant $\Delta\varepsilon_{IR}$ shift all the modes in the same IR by an equal amount.
Hence, if a crossing contains multiple instances of the same IR, $\hat H_c$ cannot break the degeneracy.
This happens, for example, at the $\Gamma$ and $M$ points of the vectorized model shown in Fig.~\ref{fig:ff_comparison}(d-f), where both HSPs contain 8 modes.
Looking at the IR composition of the modes, we find that they both have two doublets $E$, which are degenerate, in addition to four singlets.

\begin{figure}
    \centering
    \includegraphics[width=0.5\linewidth]{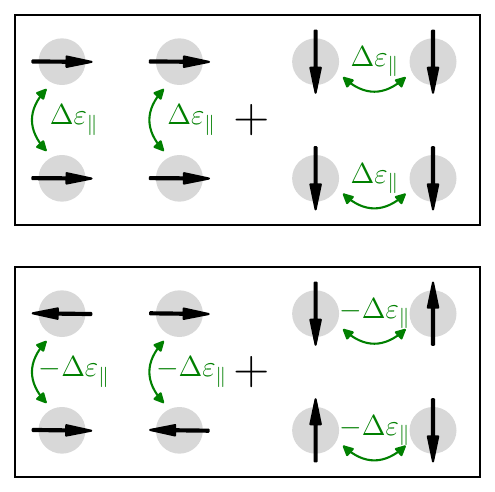}
    \caption{
    Two $E$ doublets from a 2x2 system with parallel coupling highlighted in green.
    These have the same symmetry properties in $\mathbf{C}_{4v}$, hence they cannot be separated by a constant $\Delta\varepsilon_{E}$.
    Note that when evaluating the symmetry properties, both components of the doublet need to be considered together.
    For example, in both doublets, one of the components is even under reflection $\sigma_x$ and the other is odd, hence the total effect is $1 +(- 1) = 0$, which is the corresponding character in Table \ref{tab:char_c4v}.
    }
    \label{fig:two_E_modes_suppl}
\end{figure}

One way to lift the degeneracy between these doublets is to introduce a tangential coupling between the dipoles similar to \cite{heilmann2022quasi,salerno2022loss}. As the coupling Hamiltonian $H_c$ in our approach is momentum-independent, it has the same form in momentum and real space, and thus we can use real-space considerations for designing the lifting of degeneracy.
As is shown in a simplified, but analogous 2x2 system in Fig.~\ref{fig:two_E_modes_suppl}, the two $E$ doublets have same symmetry properties in $\mathbf{C}_{4v}$, but they can be separated by coupling the $x$-components of adjacent sites in $y$-direction and the $y$-components of sites adjacent in $x$.

Collecting couplings between $x$ components into $\hat H_x$ and the couplings between $y$ components into $\hat H_y$, we can express the parallel coupling Hamiltonian in $xy$-basis as
\begin{equation}
    \hat H_{\parallel,xy} = \begin{bmatrix}
        \hat H_x & 0 \\ 0 & \hat H_y
    \end{bmatrix}.
\end{equation}
However, in the text we are working in the left-right basis, meaning that we need to take
\begin{equation}
    \hat H_{\parallel,LR} = U\hat H_{\parallel,xy}U^{-1}, \hspace{0.2cm} U = \frac{1}{\sqrt{2}}\begin{bmatrix}
        1 & i\\1 & -i
    \end{bmatrix} \otimes I_{N^2},
\end{equation}
where $I_{N^2}$ is an identity matrix matching the number of particles in the unit cell.

As the goal of this Hamiltonian is to split the $E$ doublets, and nothing else, we further project it onto the $E$ subspace of the Hamiltonian with the projection matrix $P_E$.
Thus 
\begin{equation*}
    \hat H_\parallel = P_E\hat H_{\parallel, LR}.
\end{equation*}
Because of this projection, the resulting matrix will not affect any of the other IRs, allowing us to accurately control them with their respective $\Delta \varepsilon_{IR}$.

\subsection*{Topological edge states in a semi-infinite system}

\begin{figure*}
    \centering
    \includegraphics[width=0.95\textwidth]{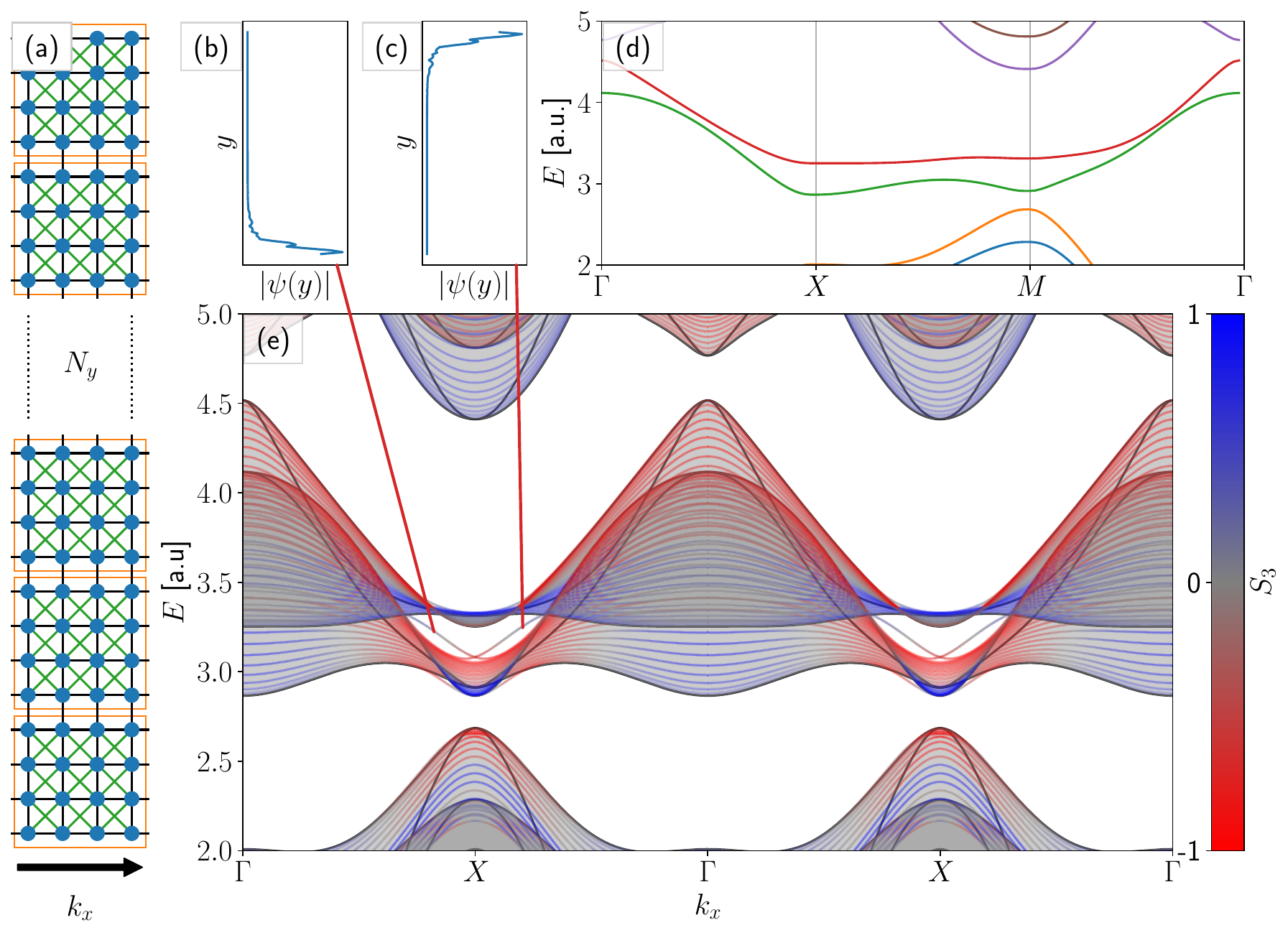}
    \caption{(a) A depiction of the configuration with only nearest and next-nearest couplings shown for clarity.
    The structure contains 30 unit-cells in the $y$-direction and periodic boundary conditions along $x$.
    (b,c) The modes crossing the gap, as highlighted in the dispersion (e), can be seen to be localized on the edges of the sample.
    (d) The energy bands of an infinite system.
    The two bands of interest (green and red) are topologically equivalent to the system discussed in Fig. \eqref{fig:ff_comparison} (c/f), with the couplings exaggerated to show better the modes crossing the gap.
    (e) The energy eigenvalues of the semi-infinite Hamiltonian with the color indicating the spin-polarization of the band.
    The shaded regions show the bulk bands projected on the $k_x$ axis.
    The full energies and eigenvectors are available at \cite{full_data}.
    }
    \label{fig:ff_comparison2}
\end{figure*}

We can further demonstrate the topology of the model by constructing a system with periodic boundary conditions along the $x$ direction, and open boundaries along $y$.
For parameters of Fig.~\ref{fig:ff_comparison}(c), the absence of a direct band gap would make the identification of the edge-states more difficult. 
To enhance their visibility, we exaggerate the couplings in $\hat{H}_c$, with $\Delta \varepsilon_{A_1} = -2.2, \Delta \varepsilon_{A2} = -1.8, \Delta \varepsilon_{B1} = 10, \Delta \varepsilon_{B2} = 10, \Delta \varepsilon_E = 1$ with additional parallel coupling of $\Delta \varepsilon_\parallel = -1.5$. 
These adjustments are carefully done so as not to generate new crossings for the third and fourth bands, so that Fig.~\ref{fig:ff_comparison}(c) and Fig.~\ref{fig:ff_comparison2} are topologically equivalent.

By calculating the eigenmodes as a function of $k_x$, with 30 unit cells along the $y$-direction, we identify two modes crossing the band gap.
Looking at the magnitude of the eigenvector as a function of the $y$-coordinate, we can verify that these modes belong to the chiral edge-states dictated by the bulk-boundary correspondence.
As the band in the (c/f) configuration of Fig.~\ref{fig:ff_comparison} is topologically equivalent to that of Fig.~\ref{fig:ff_comparison2}, these edge-states also exist for that system -- however, due to the lack of a direct band gap, they are located inside the continuum of modes.
The coloring in Fig.~\ref{fig:ff_comparison2} shows the spin-polarization of the mode in the far-field i.e. $S_3 = (|E_L| - |E_R|)/(|E_L| + |E_R|)$.
The bands display non-trivial spin polarizations throughout, but interestingly, the observed edge-states are almost purely linear.

\section*{Acknowledgments}K.A. acknowledges financial support from the Vilho, Yrjö and Kalle Väisälä Foundation of the Finnish Academy of Science and Letters. 
G.S. was funded by the Research Council of Finland through project n. 13354165 and by the Italian Ministry of University and Research under the Rita Levi-Montalcini program. 
This work was supported by the Research Council of Finland under Project Nos.~349313 and 318937 (PROFI) and by the Jane and Aatos Erkko Foundation and the Technology Industries of Finland Centennial Foundation as part of the Future Makers funding program.

\section*{Author contributions}
P.T. initiated and supervised the project. K.A. developed the theoretical model and performed numerical simulations, with input from G.S., who also contributed to the supervision of the work. All authors discussed the results and wrote the manuscript together.

\section*{Competing Interests}
The authors declare no competing interests.

\bibliography{lib}
\bibliographystyle{unsrt}

\end{document}